\documentstyle[preprint,aps,eqsecnum,psfig]{revtex}
\tightenlines
\def\m@thcombine#1#2{%
  \setbox0=\hbox{$#1$}
  \setbox1=\hbox{$#2$}
  \ifdim\wd0>\wd1
    \setbox0=\hbox to\wd1{\hss\box0\hss}
  \else
    \setbox1=\hbox to\wd0{\hss\box1\hss}
  \fi
  \mathop{\vcenter{
    \offinterlineskip\box0\box1}}}
\def\lesim{\m@thcombine<\sim}
\def\gesim{\m@thcombine>\sim}

\begin{document}

\draft
\title{ A GAUGE INVARIANT FORMULATION OF INTRINSICALLY NONPERTURBATIVE QCD }

\author{V. Gogohia }

\address{HAS, CRIP, RMKI, Theory Division, P.O.B. 49, H-1525 Budapest 114, Hungary \\
 email addresses: gogohia@rmki.kfki.hu \ and \ gogohia@rcnp.osaka-u.ac.jp}
%Research Center for Nuclear Physics (RCNP), Osaka University \\
%          Mihogaoka 10-1, Ibaraki, Osaka 567-0047, Japan  \\ }

\maketitle

\begin{abstract}
Using a system of the corresponding Schwinger-Dyson equations of motion, a     
pure dynamical theory of quark confinement and spontaneous breakdown of        
chiral symmetry is formulated. It is based on dominated in the QCD vacuum      
self-interaction of massless gluons only, i.e., without involving some extra   
degrees of freedom. This interaction becomes strongly singular in the deep     
infrared domain leading thus to the enhancement of zero momentum modes in the  
nonperturbative QCD vacuum. Using theory of distributions, complemented by     
the dimensional regularization method, we have explicitly shown that strong    
infrared singularities can be put under control. In this  
way a new phase in QCD, intrinsically nonperturbative QCD which is        
manifestly gauge-invariant, was discovered.  
As a result, a highly nontrivial dynamical and topological structure of the    
QCD vacuum has emerged within our approach.  
We have also explicitly shown how infrared multiplicative renormalization program should be done in order to self-consistently remove all the strong infrared 
singularities from the theory. The corresponding convergence conditions play   
a crucial role in this program.  
In this theory any physical observables are determined by such correlation     
functions from which all types of the perturbative contributions should be     
subtracted, by definition. Theory is not only infrared finite but it is free
from the ultraviolet divergences as well. It has a mass gap $\Delta >0$, i.e., 
there are no physical states in the interval $(0, \Delta)$. 
It explains confinement, spontaneous breakdown of chiral         
symmetry, and other nonperturbative effects on a general ground and in         
self-consistent way.
\end{abstract}


\vfill
\eject

\section{Introduction}

The full dynamical information of any quantum gauge field theory such as       
quantum chromodynamics (QCD) is contained in the
corresponding quantum equations of motion, the so-called
Schwinger-Dyson (SD) equations for lower (propagators) and higher
(vertices and kernels) Green's functions [1-3]. It is a highly nonlinear,      
strongly coupled system of four-dimensional integral
equations for the above-mentioned quantities. The Bethe-Salpeter (BS) type     
integral equations for the bound-state amplitudes [4] should be also included  
into this system. The kernels of
these integral equations are determined by the infinite series of
the corresponding skeleton diagrams. It is a general
feature of $nonlinear$ systems that the number of exact solutions
(if any) can not be fixed $a \ priori$. Thus formally it
may have several exact solutions. These equations should be
complemented by the corresponding Slavnov-Taylor (ST)
identities [5] which in general relate the above mentioned
lower and higher Green's functions to each other. These
identities are consequences of the exact gauge invariance and
therefore $"are \ exact \ constraints \ on \ any \ solution \ to
\ QCD"$ [1]. Precisely this system of equations can serve as an
adequate and effective tool for the nonperturbative (NP) approach to
QCD [6]. To say today that QCD is the NP theory is almost tautology.           
The problem is  
how to define it exactly since we surely know that QCD has a perturbative (PT)
phase as well because of asymptotic freedom (AF) [7].

\section{Yang-Mills sector}

In order to exactly define NP QCD, let us start first with the Yang-Mills (YM) 
sector. The two-point Green's function, describing the full gluon    
propagator, is

\begin{equation}
iD_{\mu\nu}(q) = \left\{ T_{\mu\nu}(q)d(-q^2, \xi) + \xi L_{\mu\nu}(q) \right\} {1 \over q^2 }.
\end{equation}
Here $\xi$  is a gauge fixing parameter ($\xi = 0$, Landau gauge) and         
$T_{\mu\nu}(q)=g_{\mu\nu}-(q_{\mu} q_{\nu} / q^2) = g_{\mu\nu } - L_{\mu\nu}(q)$.
Evidently, its free (tree level) counterpart is
obtained by simply setting the full gluon form factor $d(-q^2,
\xi)=1$ in Eq. (2.1) (see Eq. (2.7) below). The solutions of the
above-mentioned SD equation for the full gluon propagator (2.1)
are supposed to reflect the complexity of the quantum structure
of the QCD ground state. Precisely this determines the central role of the full
gluon propagator in the SD system of equations [8]. The   
SD equation for the full gluon propagator is a highly nonlinear system of      
four-dimensional integrals containing many different, unknown in general       
propagators, vertices and kernels, so there is no hope for exact solution(s).  
However, in
any case the solutions of this equation can be distinguished from each other by
their behavior in the deep infrared (DIR) limit, describing thus many
(several) different types of quantum excitations and fluctuations of gluon
field configurations in the QCD vacuum. The ultraviolet (UV)    
limit of these solutions is uniquely determined by AF.
Evidently, not all of them can reflect the real structure of the QCD vacuum.  

  The DIR asymptotics of the full gluon propagator can be generally classified 
into the two different types: singular which means the IR enhanced (IRE) or   
smooth which means the IR finite (IRF) or even the IR vanishing (IRV) gluon    
propagators. However, any deviation in the behavior of the full
gluon propagator in the DIR domain from the free one
automatically assumes its dependence on a scale parameter (at
least one) in general different from QCD asymptotic scale
parameter $\Lambda_{QCD}$. It can be considered as responsible
for the NP dynamics (in the IR region) in the QCD
vacuum. If QCD itself is a confining theory, then   
such characteristic scale is very likely to exist. In what follows,
let us denote it, say, $\mu$. This is very similar to AF
which requires the above-mentioned asymptotic
scale parameter associated with nontrivial PT dynamics
in the UV region (scale violation). In this connection it is worth emphasizing 
that being numerically a few hundred $MeV$ only, it cannot survive in the UV   
limit. This means that none of the finite scale parameters, in particular      
$\Lambda_{QCD}$, can be determined by PT QCD. It should come from the 
IR region, so it is NP by origin. How to establish the relation between these  
two independent scale parameters was shown in our paper [9]. Despite the PT    
vacuum cannot be the QCD true ground state [10], nevertheless, the existence   
of such kind of  
the relation is a manifestation that "the problems encountered in perturbation 
theory are not mere mathematical artifacts but rather signify deep properties  
of the full theory" [11]. The message that we have tried to convey is that     
precisely AF clearly indicates the existence of the NP phase in the full QCD.

\subsection{Truly NP phase}

The phenomenon of ''dimensional    
transmutation'' [1,12] only supports our general conclusion that QCD exhibits  
a mass determining the scale of NP dynamics in its ground state. In terms    
of the full gluon propagator it can be exactly defined as follows:             
                                 
\begin{equation}
d^{TNP}(-q^2, \mu^2) = d(-q^2, \mu^2) - d(-q^2, \mu^2 =0),
\end{equation}
where we introduce explicit dependence on the NP mass scale $\mu$, and the    
dependence on the gauge fixing parameter as well as other possible          
parameteres is  
omitted for simplicity. Thus this subtraction can be also considered as the    
definition of the truly NP (TNP) part
of the full gluon propagator since when the NP scale parameter goes formally   
to zero then the TNP part vanishes, 
i.e., it uniquely determines the TNP part of the full gluon propagator.        
On the other hand, the definition (2.2) explains the difference between the 
TNP part $d^{TNP}(-q^2)$ and the full gluon propagator
$d(-q^2)$ which is NP itself.                                                  
Let us note now that the
limit $\mu^2 \rightarrow 0$ is usually equivalent to the
limit $-q^2 \rightarrow \infty$. In some other cases, the gluon
propagator does not depend explicitly on the NP
scale parameter (the dependence is hidden) then its behavior at
infinity should be subtracted in Eq. (2.2). 

It is almost obvious that because of AF one can identify

\begin{equation}
d(-q^2, \mu^2 =0) \equiv d^{PT}(-q^2)
\end{equation}
since any Green's function in particular the gluon one should approach its PT  
counterpart (up to PT renormgroup log improvements) in the UV limit 
($\mu^2 \rightarrow 0$ which is equivalent to $-q^2 \rightarrow \infty$,  
as was emphasized above). It is clear that by "PT" we mean intermediate 
(IM) plus UV regions (since the IM region still remains a "$terra \ incognita$"
in QCD, though it seems to us that it can be sufficiently well approximated by 
the free gluon propagator, see Eq. (2.7) below). Obviously, the relation (2.3) 
is our definition of the PT phase in QCD. Thus the relation (2.2) becomes
 
\begin{equation}
d(-q^2, \mu^2) = d^{TNP}(-q^2, \mu^2) + d^{PT}(-q^2).
\end{equation}
Let us underline that this is an exact relation (the gluon momentum runs
over the whole range $[0, \infty)$), so at this stage there is no      
approximation made. Due to the inevitable identification (2.3), one can say 
that we define the NP scale parameter in the way that when it formally goes    
to zero (i.e., when the TNP part
vanishes) then the PT phase only survives in the full gluon propagator.
This also allows one to clearly distinguish the NP scale        
parameter between all other mass scale parameters which may be presented in the
full gluon propagator. In some special cases, it can be identified with the    
mass of the particle. For example, in the dual Abelian Higgs model the mass of 
the dual gauge boson serves as the NP scale [13].
 It is worth emphasizing that in the realistic   
models  of the full gluon propagator its TNP part usually
coincides with its DIR asymptotics determining thus the strong
intrinsic influence of the IR properties of the theory on its
NP dynamics.

Substituting this exact decomposition into the full gluon propagator
(2.1), one obtains

\begin{equation}
D_{\mu\nu}(q, \mu) = D^{TNP}_{\mu\nu}(q, \mu) + D^{PT}_{\mu\nu}(q),
\end{equation}
where                                       

\begin{equation}
 D^{TNP}_{\mu\nu}(q,\mu) = - i T_{\mu\nu}(q) d^{TNP}(-q^2, \mu^2){ 1 \over q^2}, 
\end{equation}

\begin{equation}
D^0_{\mu\nu}(q) = - i \{ T_{\mu\nu}(q) + \xi L_{\mu\nu}(q) \} { 1 \over q^2},
\end{equation}

\begin{equation}
D^{PT}_{\mu\nu}(q) = - i \{ T_{\mu\nu}(q)  d^{PT}(-q^2) + \xi L_{\mu\nu}(q) \}
{ 1 \over q^2}.
\end{equation}
For further purposes the explicit expression for the free gluon          
propagator $D^0_{\mu\nu}(q)$ is also given.  
Thus the exact decomposition (2.5), complemented by the definitions            
(2.6) and (2.8),   
has a few remarkable features. First of all, all the dependence of the full    
gluon propagator on the NP scale parameter $\mu$ is exactly placed in its    
TNP part which vanishes as it formally goes to zero, i.e., when the PT phase  
only survives in QCD. Secondly, the explicit gauge dependence of the full      
gluon propagator is exactly shifted from its TNP part to its PT part (for      
reasons to proceed in this way, see discussion below). Thus we clearly separate
the NP phase from the PT one in QCD.

\subsection{Intrinsically NP phase}

Up to this moment we have delt only with exact decompositions and definitions, 
i.e., only algebraic manipulations have been performed. It is the time now 
to introduce nontrivial dynamics into this scheme. Evidently, the only place  
where it can be done is, of course, the DIR region, i.e., by saying something  
nontrivial about the DIR asymptotics of the full gluon propagator one can      
additionally  distinguish between its TNP and PT parts. The UV 
behavior of both parts, i.e., of the full gluon propagator, is controlled by   
AF. Fortunately, we have  
an exact criterion for establishing the DIR structure of the full gluon        
propagator.
The PT phase in QCD begins, of course, from the free gluon propagator (2.7)    
which has an exact $(1/q^2)$ singularity in the DIR limit since it is          
defined in the whole range. The PT part, presented by Eq. (2.8) 
also in the whole range, possesses the same property.  This means that the  
free gluon singularity $(1/q^2)$ at $q^2 \rightarrow 0$ is an exact separation 
line between the 
TNP and PT parts in the full gluon propagator (2.5). The PT phase in QCD is defined as one which at maximum is singular in the IR as free gluon propagator.   
Its existance in QCD is important from conceptual point of view and it is      
determined by AF in the UV limit. In what follows this type of singularities   
will be called as PT IR singularities.  

Contrary to the PT part, the TNP part of the full gluon propagator (2.6),      
where the momentum also runs over the whole range,  
should have then singularities in the DIR domain stronger than $(1/q^2)$. Of   
course,
such strong singularities can be only of dynamical origin. The only dynamical 
mechanism in QCD which can produce such severe singularities in the vacuum is  
self-interaction of massless gluons in the DIR domain.                    
Precisely this self-interaction in the UV limit leads to AF. Thus one comes to 
the inevitable conclusion that the TNP part of the full gluon propagator should
have strong IR singularities different from those of the PT part.              
In what follows this type of singularities will be called as
NP IR singularities. Thus they should be summarized by the full gluon propagator and effectively correctly described by its TNP part in the DIR domain. In
this way the TNP becomes intrinsically NP (INP) one. So the definition of 
the INP YM quantum theory consisits of the two conditions.  

I). The first necessary condition defines the TNP part of the full gluon       
propagator (2.6), on account of the definition (2.2).

II). The second sufficient condition specifies the existence and structure of  
more stronger than $(1/q^2)$ IR singularities in the QCD NP vacuum.

Now it becomes transparently clear why the gauge fixing parameter $\xi$ should 
be shifted to the PT part of the full gluon propagator. Its
longitudinal part is exactly singular in the IR as $(1/q^2)$.
So, by definition, it  belongs to the PT world and therefore it does not make  
any sense to decompose $\xi$ similar to Eq. (2.2) though formally it is        
possible to do, of course.    
Thus the INP part of the full gluon propagator is manifestly gauge invariant   
and only transfer (physical) degrees of freedom of the gauge bosons are        
important for the NP dynamics in QCD.

The INP YM theory will become INP QCD when we will include into this       
scheme quark, ghost and quark-ghost sectors as well (see below). However,      
beforehand, we have to characterize $quantitatively$ the QCD NP vacuum which   
arises in our approach.

\section{ZMME quantum model of the QCD ground state}

The quantum structure of the QCD NP vacuum is dominated by such types of       
excitations and fluctuations of gluon field configurations there which are due 
to self-interaction of massless gluons since precisely this interaction is 
the main quantum, dynamical effect in QCD.\footnote{Classical field            
configurations may also exist in the QCD NP vacuum since its dynamical and topological structure can be organized at different levels: quantum and classical. 
However, they are by no means dominant.} 
In the UV region it implies AF. In the DIR region it becomes strongly singular 
and thus  
becomes responsible for the enhancement of zero momentum modes in the QCD NP   
vacuum. All the NP IR singularities should be absorbed into full gluon         
propagator while all other Green's functions can be regarded as regular        
functions with   
respect to the momentum transfer, i.e., gluon momentum $q$ (see next section). 
Thus that is the zero momentum modes are enhanced in the QCD NP vacuum is in
complete agreement with our conclusion driven in the preceding section. They  
are behind the existence of the INP phase in QCD.    
    
Our quantum, dynamical model of the QCD true ground state is based on the 
existence and importance of such kind of the NP excitations and fluctuations   
of gluon field configurations which are due to self-interaction of massless gluons only without explicit involving some extra degrees of freedom. They   
are to be summarized by the INP part of the full gluon propagator and are to be
effectively correctly described by its behavior in the DIR domain. As was   
explained above, these types of gluon field configurations necessary           
correspond to the strongly singular in the IR gluon propagator.  
 In what follows we will call our model of the QCD ground   
state as zero momentum modes enhancement (ZMME) or simply zero modes           
enhancement (ZME, since we work always in the momentum space) quantum model  
[14,15].

Before going into the brief descriptoin of the distribution nature of the NP   
IR singularities, introduced in Eq. (3.1) below, a few short remarks are in    
order. From QCD sum rules [16] it is well known that AF is stopped by          
power-type terms reflecting the growth of the coupling in the IR. Approaching  
the DIR region from the above, the IR sensitive contributions were parametrized
in terms of a few quantities (gluon and quark condensates, etc.) while direct  
access to NP effects (i.e., to the DIR region) was blocked by the IR           
divergences [16,17]. Our approach  
to NP QCD, in particular its true ground state, is a further step into the
DIR region (in fact, we are deeply inside it) since the distribution theory    
(DT) allows one to correctly deal with the NP IR singularities (see below).

In general, all the Green's functions in QCD are generalized functions, i.e.,  
they are distributions [18]. Especially this is true for the severe NP IR      
singularities    
due to self-interaction of massless gluons in the QCD vacuum. Roughly speeking,
 this means that all relations involving distributions should be considered   
under corresponding integrals taking into account smoothness properties of the 
corresponding test functions. Not loosing generality, the severe NP IR         
singularities can be analytically taken into account  
in terms of the TNP gluon form factor in Eq. (2.6) with a Euclidean signature  
($-q^2 \rightarrow q^2$) as follows:

\begin{equation}
d^{INP}(q^2, \mu^2)= (\mu^2)^{-\lambda - 1} (q^2)^{\lambda} \times f(q^2),
\end{equation}
where obviously we include $1/q^2$ from Eq. (2.6) into the exponent $\lambda$
which in general is arbitrary (any complex number with $Re \lambda < 0$).
The function $f(q^2)$ is a dimensionless function  
which is regular at zero and otherwise remaining arbitrary, but preserving 
AF in the UV limit. This is nothing else but the analytical         
formulation of the second sufficient condition of the existence of the INP     
phase in the YM theory. That is why we replaced
superscipt ``TNP'' by ``INP'' in Eq. (3.1).
Since we are particulary interested in the DIR region, the arbitrary function 
$f(q^2)$ should be also expanded   
around zero in the form of the Taylor series in powers of $q^2$, i.e,          
     
\begin{equation}
f(q^2) = \sum_{m=0}^{[- \lambda] -1} (q^2)^m f^{(m)}(0) +                      
\sum_{m=[-\lambda]}^{\infty} (q^2)^m f^{(m)}(0),  
\end{equation}
where $[-\lambda]$ denotes its integer number and

\begin{equation}
f^{(m)}(0) = { 1 \over m!} \Biggl( {d^m f(q^2) \over d(q^2)^m} \Biggr)_{q^2=0}.
\end{equation}
As a result, we will be left with finite sum  
of power terms with exponent decreasing by one starting from $- \lambda$. All
other remaining terms from the Taylor expansion (3.2) starting from the term   
having already the PT IR singularity (the second sum in Eq. (3.2)) should be   
shifted to the PT part of the full gluon propagator.                           
                                         
Let us consider now for simplicity only, the most strong NP IR     
singularity presented by
the first term in the above-mentioned Taylor expansion. DT, elaborated and     
developed by Gel'fand and Shilov in Ref. [19], tells us  
that though the exponent $\lambda$ in general is arbitrary        
the distribution $(q^2)^{\lambda}$ will have a simple pole at points  
$\lambda = - (n/2) - k, \ (k=0, 1, 2, 3...)$, where $n$ denotes the number    
of dimensions in Euclidean space ($q^2 = q^2_0 + q^2_1 + q^2_2 + ... + q^2_{n-1}$). In order to actually define the system of SD equations in the DIR
domain, it is necessary to introduce the IR regularization parameter   
$\epsilon$, defined as $D = n + 2 \epsilon, \ \epsilon \rightarrow 0^+$       
within a gauge invariant dimensional regularization (DR) method of 't Hooft   
and Veltman [20]. As a result, all the Green's functions should be regularized 
with respect to $\epsilon$ (see next section), which is to be set to zero at   
the end of computations. The structure of the NP IR singularities is then      
determined (when $D$ and consequently $n$ are even numbers) as follows [19]:   
                                                    
\begin{equation}
(q^2)^{\lambda} = { C_{-1}^{(k)} \over \lambda +(D/2) + k} + finite \ terms,
\end{equation}
where the residue is 

\begin{equation}
 C_{-1}^{(k)} = { \pi^{n/2} \over 2^{2k} k! \Gamma ((n/2) + k) } \times        
L^k \delta^n (q)
\end{equation}
with                                                                           
 
\begin{equation}
L = {\partial^2 \over \partial q^2_0} + {\partial^2 \over \partial q^2_1}
+ ... + {\partial^2 \over \partial q^2_{n-1}}. 
\end{equation}

 Let us underline the most remarkable features of these expressions. The   
order of singularity does not depend on $\lambda$, $n$ and $k$, i.e.,
it is always a simple pole, $(1/ \epsilon)$, in terms of the IR regularization 
parameter $\epsilon$. This means that all power terms in Eq. (3.1) will have  
the same singularity, i.e.,                                         

\begin{equation}
(q^2)^{- {D \over 2} - k + \epsilon} = { 1 \over \epsilon} C_{-1}^{(k)} +      
finite \ terms,
\end{equation}
where we can put $D=n$ now. However, the residue at pole will be 
drastically changed from one power singularity to another. This means different
solutions to the whole system of SD equations for different set of numbers    
$\lambda$ and $k$. Different solutions    
means in turn different vacua. Thus in our picture different vacua are to be   
labelled by two numbers: the exponent $\lambda$ and $k$, or equivalently,    
$D(=n)$ and $k$. At given number of $D(=n)$ the exponent  
$\lambda$ is always negative being integer if $D(=n)$ is even number or        
fractional if $D(=n)$ is odd number. The number $k$ is always integer and      
positive and precisely it determines the corresponding residue at pole, see
Eq. (3.5). At given number of $k$, apart from unimportant      
finite constants, the residue at any $D(=n)$ is the same. This means in turn   
the infinite number of the same vacua, i.e., ``degenerate'' vacua with 
differences due to concrete numerical value of $D(=n)$ only. On the other hand,
at given $D(=n)$ and different $k$, the infinite number of vacua will be       
drastically different. Let us remind the reader that we also have the finite   
number of  
terms which number is determined by the exponent $\lambda$ itself, i.e., how   
many terms should be kept in the Taylor expansion of $f(q^2)$ function in      
Eq. (3.2). It would be not surprised if these numbers were 
somehow related to the nontrivial topology of the QCD NP vacuum.      

Thus the wide-spread opinion that severe IR singularities in QCD,         
stronger than $(1/q^2)$, cannot be controlled is not justified. Precisely DT   
[19], complemented by the DR method [20], enables one to    
correctly treat them at any arbitrary exponent $\lambda$. The structure of    
the NP IR singularities with a Minkowski signature is much more complicated
since due to light cone kinematical singularities also appear. For odd number  
of $D(=n)$ the poles of second order, i.e., $(1/ \epsilon)^2$, will appear even
in Euclidean space [19]. For correctly
regularized Green's functions, their Fourier transform exists, of course, and 
some of them can be found in Ref. [19]. However, the structure of the NP IR singularities in configuration space is also rather complicated. That is why we   
prefer to work always in momentum space with a Euclidean signature.    
At least, it guarantees that all the NP IR singularities are of dynamical      
origin. 

 There exists a principle difference between the NP IR divergences 
and PT UV ones which can be regularized within the DR method as               
$D = n - 2 \delta, \ \delta \rightarrow 0^+$.
The former are always simple pole, $1 / \epsilon$,
not depending on the exponent $\lambda$ which roughly speaking can be related 
to a number of closed loops. The PT UV divergences heavily depend on it so that
they become $(1 / \delta)^{\lambda}$, i.e., higher order poles necessarily     
appear in multi-loop diagrams.  

Concluding general discussion of the structure of the NP IR singularities in   
QCD, let us explicitly demonstrate the instructive correspondence between     
them in four dimensional (4D) and two dimensional (2D) QCD.   

A). In 4D QCD the simplest NP IR singularity is determined by set of numbers   
which label the vacuum in this case as follows: $\lambda = - 2$ and $k=0$. So 
the corresponding distribution (3.7) becomes  
                                 
\begin{equation}
(q^2)^{-2 + \epsilon} = {\pi^2  \over \epsilon } \delta^4 (q) + finite \ terms.
\end{equation}

B). In 2D QCD [21], the set of numbers which label the simplest vacuum is:     
$\lambda =-1$ and $k=0$. So again from Eq. (3.7), one obtains
 
\begin{equation}
(q^2)^{-1 + \epsilon} = {\pi \over \epsilon } \delta^2 (q) + finite \ terms.   
\end{equation}
Thus 2D QCD with $(q^2)^{-1}$ gluon propagator (samultaneously it is the    
simplest NP IR singularity which underlines the special status of 2D QCD)
has the same NP IR structure of singularities as 4D QCD has       
with $(q^2)^{-2}$ behavior of the gluon propagator in DIR and vice versa.  
But we know that 2D QCD confines quarks via the linear rising potential, at    
least in the large $N_c$ limit [21].  
This exact similarity provides that 4D QCD also should confine quarks (see     
below). The structure of 4D QCD NP vacuum, however, much more
complicated than in 2D QCD and 3D compact QED, where color confinement does indeed take place [22]. In the former case confinement can be analysed in 
terms of the linear rising potential only for heavy quarks when planar approximation like in 2D QCD becomes relevant for the one gluon exchange diagram with  
point-like vertices within the Wilson loop approach [23].

Let us underline, however, the principal difference between    
2D QCD and 4D QCD. As was mentioned above, in 2D QCD the free gluon singularity  $(q^2)^{-1}$ is the simplest NP IR singularity at the same time. 
This means, in turn, that in 2D QCD there is no possibility to additionaly separate the INP phase from TNP one. Moreover, even how to define the TNP phase also becomes problematic since 
in 2D QCD the coupling constant has dimension of a mass, i.e, in fact  
the subtraction in Eq. (2.2) becomes trivial in this theory. However, it is possible to explicitly show that the nonperturbative 2D QCD can be also formulated
in manifestly gauge invariant way (see our recent hep-ph/0104296 preprint as   
well as Appendix below).

\section{Quark sector}

Along with the full gluon propagator, the quark propagator also plays one of   
the dominant roles in QCD. The quark Green's function satisfies its own SD     
equation, namely 

\begin{equation}
S^{-1} (p) = S_0^{-1} (p)- g^2_F \int {d^nq \over (2 \pi)^n} \Gamma_{\mu}      
(p,q) S(p-q) \gamma_{\nu} D_{\mu\nu}(q),
\end{equation}
where $g^2_F = g^2 C_F$ and $C_F$ is the eigenvalue of the quadratic Casimir   
operator in the fundamental representation. $\Gamma_{\mu} (p, q)$ is the corresponding quark-gluon proper vertex.
The free quark propagator is $S_0^{-1} (p) = - i(\hat p - m_0)$
with $m_0$ being the current (``bare'') quark mass. 
Substituting the general, exact decomposition (2.5) into the quark SD equation
(4.1), one obtains (still yet Minkowski metric)

\begin{equation}
S^{-1} (p) = S_0^{-1} (p)+ \Sigma^{TNP} (p) + \Sigma^{PT} (p),
\end{equation}
where

\begin{equation}
\Sigma^{TNP} (p) = - g^2_F \int {d^nq \over (2 \pi)^n} \Gamma_{\mu} (p, q)     
S(p-q) \gamma_{\nu} D_{\mu\nu}^{TNP}(q, \mu),
\end{equation}
and $D_{\mu\nu}^{TNP}(q, \mu)$ is given in Eq. (2.6).

\begin{equation}
\Sigma^{PT} (p) = - g^2_F \int {d^nq \over (2 \pi)^n} \Gamma_{\mu} (p, q) S(p-q) \gamma_{\nu} D_{\mu\nu}^{PT}(q),
\end{equation}
and $D_{\mu\nu}^{PT}(q)$ is given in Eq. (2.8). Obviously, $\Sigma^{TNP} (p)$  
and $\Sigma^{PT} (p)$ are the TNP and PT contributions into the quark          
self-energy, respectively. Thus this equation determines the TNP quark SD     
equation in NP QCD.

\subsection{INP quark SD equation}

In order to go further to determine the INP quark SD equation it is necessary
the TNP part of the full gluon propagator, 
$D_{\mu\nu}^{TNP}(q, \mu)$, to replace by its INP counterpart, i.e.,           
to substitute 
$d^{INP}(q^2, \mu^2)$ given in Eq. (3.1) instead of  $d^{TNP}(q^2, \mu^2)$ in  
Eq. (2.6). The previous quark SD equation 
(4.2)-(4.4) thus becomes (Euclidean metric is already assumed)  

\begin{equation}
S^{-1} (p) = S_0^{-1} (p)+ \Sigma^{INP} (p) + \Sigma^{PT} (p),
\end{equation}
where the PT part remains unchanged, of course, and

\begin{equation}
\Sigma^{INP} (p) =  - g^2_F i \int {d^nq \over (2 \pi)^n} \Gamma_{\mu} (p, q) 
S(p-q) \gamma_{\nu} D_{\mu\nu}^{INP}(q, \mu).
\end{equation}                            
 
However, the important problem immediately arises. Confining to YM sector only,
it was inevitable to conclude that all the 
NP IR singularities could be absorbed by the full gluon propagator. In the full
QCD additional sourses of the NP IR singularities may appear. For       
example, the proper vertex $\Gamma_{\mu} (p, q)$ in this quark SD equation can,
in principle, suffer from the NP IR singularities as well.       
Not loosing generality then it can be presented, similar to Eq. (3.1), as     
follows:                                                                       

\begin{equation}
\Gamma_{\mu} (p, q) = (q^2)^{-\alpha} \bar \Gamma_{\mu} (p, q), \quad          
Re \alpha > 0,             
\end{equation}
where $\bar \Gamma_{\mu} (p, q)$ is already regular with respect to $q$ and is 
of the corresponding dimension. However, the PT part of the quark
SD equation (4.5) becomes too singular now. All the terms violating the        
definition of the PT part to be in the DIR limit as singular as $(1/q^2)$, by  
expanding $\bar \Gamma_{\mu} (p, q)$ around zero in form of the Taylor series  
in powers of $q^2$, should be shifted to the INP part of the quark SD equation 
(4.6), i.e.,

\begin{equation}
\bar \Gamma_{\mu} (p, q) = \sum_{m=0}^{[\alpha]-1}(q^2)^m \bar \Gamma_{\mu}^{(m)}
(p, 0) + \sum_{m=[\alpha]}^{\infty}(q^2)^m \bar \Gamma_{\mu}^{(m)}(p, 0),      
\end{equation}
where $[\alpha]$ denotes its integer number and 

\begin{equation}
\bar \Gamma_{\mu}^{(m)} (p, 0) = { 1 \over m!} \Biggl( {\partial^m \over \partial(q^2)^m } \bar \Gamma_{\mu}^{(m)} (p,q) \Biggr)_{q^2=0}
\end{equation}
Evidently, in the INP part of the SD equation (4.6) the NP IR singularities    
due to vertex can be easily incorporated into the gluon propagator by simply   
redefining exponent in Eq. (3.1) from $\lambda$ to $\lambda - \alpha$, so this
will not cause any broblems. However, serious problem will be caused by the 
above-mentioned finite number of terms shifted from the PT part, namely the first sum in Eq. (4.8) multiplied by $(q^2)^{-\alpha}$ as well as by       
tensor factor which explicit expression is not shown
for simplicity. In the PT limit ($\mu^2 \rightarrow 0$), when only the PT   
phase survives and the INP contribution should       
vanish, these terms will remain, however. This obviously contradicts our first 
necessary condition of the definition of the INP phase in QCD. The only way to 
satisfy it is to put                                                           

\begin{equation}
\alpha=0$, so that $\sum_{m=0}^{-1} = 0,                                       
\end{equation}      
by definition. Absolutely in the same way, by using the corresponding SD       
equation for the ghost self-energy, can be shown that the ghost-gluon vertex

\begin{equation}
G_{\mu} (k, q) = k^{\lambda} G_{\mu\lambda} (k, q)             
\end{equation}
should be also regular with respect to momentum transfer $q$.
Thus within our approach to NP QCD, the higher Green's functions (quark-gluon  
vertex as well as ghost-gluon vertex) should be considered as regular ones     
with respect to momentum transfer. However, let us note in advance that in the 
quark-ghost sector, in particular in quark ST identity        
momentum transfer goes through the ghost self-energy (momentum $k$ in previous 
expression), i.e., momentum transfer  
coincides with its momentum. Due to the above-mentioned this means in turn that
the quark-gluon vertex is regular with respect to the ghost sel-energy momentum
$k$ as well. Moreover, it is possible to explicitly show that the ghost self-energy is also can be regular function of its momentum within our approach,      
i.e., it can exist and be finite at zero in our scheme.\footnote{In principle, 
a singular dependence of the ghost self-energy on its momentum should be not excluded
$a \ priori$. In this case, however, the general treatment of the NP IR singularities is completely different especially in the quark-ghost sector and therefore requires a separate consideration. Also the smoothness properties of the corresponding test functions are compromised in this case, i.e., the use of the   
relation (3.4) becomes problematic, at least in the standard DT sense.} So the 
ghost degrees of freedom are regular functions of their momenta in the present 
investigation. Especially this is true for the simplest NP IR singularities    
(at $k=0$) possible in nD QCD.            
At the same time, all the NP IR singularities in the full   
QCD vacuum due to self-interaction of massless gluons there can be summarized 
by the gluon propagator only, indeed.

\subsection{IRMR program}

 As was mentioned above,  all the Green's functions
became dependent generally on the IR regularization
parameter $\epsilon$, i.e., they became IR regularized. This dependence is     
not explicitly shown for simplicity. 
Let us introduce the IR renormalized (finite) quark-gluon vertex function,     
quark propagator and coupling constant as follows:

\begin{eqnarray}
\Gamma_\mu(p, q) &=& Z^{-1}_1(\epsilon) \bar \Gamma_\mu(p, q), \nonumber\\
S(p) &=& Z_2( \epsilon) \bar S(p), \qquad \epsilon \rightarrow  0^+ \nonumber\\g^2 &=& X(\epsilon) \bar g^2. 
\end{eqnarray}
Here $Z_1(\epsilon), \ Z_2(\epsilon)$ and $X(\epsilon)$
are the corresponding IR multiplicative renormalization (IRMR) constants.
The $\epsilon$-parameter dependence is indicated explicitly
to distinguish them from the usual UVMR constants. In all relations here and   
below containing the IRMR constants, the
$ \epsilon \rightarrow  0^+ $ limit is always assumed at final stage.
$\bar \Gamma_\mu(p, q)$ and $\bar S(p)$ are the IR renormalized (finite)       
 Green's functions and therefore do not depend on $\epsilon$ in the
$ \epsilon \rightarrow  0^+ $ limit, i.e. they exist as
$ \epsilon \rightarrow  0^+ $ as well as the IR finite coupling constant $\bar
g^2$ (charge IR renormalization).      
There are no restrictions on the
$ \epsilon \rightarrow  0^+ $ limit behavior of the IRMR
constants apart from the regular $\epsilon$ dependence of the  quark wave      
function IRMR constant $Z_2(\epsilon)$ (see Eq. (4.17) below).

As we have established in preceding sector, the NP IR singularities in the QCD 
vacuum, summarized by the DIR asymptotics of the full gluon propagator, are    
to be presented in its INP part (3.1). In terms of the IR regularization       
parameter $\epsilon$, it has always a simple pole does not depending on the   
exponent $\lambda$. This means that the INP part of the full gluon propagator
(and consequently the full gluon propagator itself) should be renormalized as
follows (Euclidean metric): 

\begin{equation}
D_{\mu\nu}^{INP}(q, \mu) = { 1 \over \epsilon} \bar D_{\mu\nu}^{INP}(q, \mu),
\qquad      \epsilon \rightarrow  0^+,
\end{equation}
and

\begin{equation}
D_{\mu\nu}(q, \mu) = Z_D \bar D_{\mu\nu} (q, \mu) = { 1 \over \epsilon} \bar D_{\mu\nu}(q, \mu), \qquad  \epsilon \rightarrow  0^+,
\end{equation}
i.e., the IRMR constant of the full gluom propagator $Z_D(\epsilon)$ is always 
$(1 / \epsilon)$ and $\bar D_{\mu\nu}^{INP}(q, \mu)$ as well as $\bar D_{\mu\nu}(q, \mu)$ exists as $\epsilon \rightarrow  0^+$. At the same time, its PT part
should be considered as IR renormalized from the very beginning, i.e.,         

\begin{equation}
D_{\mu\nu}^{PT}(q) = \bar D_{\mu\nu}^{PT}(q)                                   
\end{equation}   
since it is certainly free from the NP IR singularities, by definition. 

Substituting all these relations into the quark
SD equation (4.5), one obtains that a cancellation of the NP IR divergences    
takes place if and only if (iff) 

\begin{equation}
X(\epsilon) Z^2_2(\epsilon) Z^{-1}_1(\epsilon) = \epsilon Y_q, \qquad          
\epsilon \rightarrow 0^+
\end{equation}
holds. Here $Y_q$ is the arbitrary but finite constant. The relation      
(4.16) is a quark SD equation convergence condition in the most general form.   

The IR finite quark SD equation becomes (Euclidean version)

\begin{equation}
\bar S^{-1} (p) = Z_2(\epsilon) S_0^{-1} (p)+ \bar \Sigma^{INP} (p),
\end{equation}
where

\begin{equation}
\bar \Sigma^{INP} (p) = - \bar g^2_F Y_q i \int {d^nq \over (2 \pi)^n} \bar \Gamma_{\mu} (p, q) \bar S(p-q) \gamma_{\nu} \bar D_{\mu\nu}^{INP}(q, \mu).
\end{equation}

The remarkable feature of our IRMR program is that the contribution
from the PT part vanishes since it becomes of order $\epsilon$ as              
$\epsilon \rightarrow 0^+$ because of the convergence condition (4.16). 
Here and everywhere (ghost, quark-ghost sectors, etc.) the terms containing
the logarithm of the coupling constant (scale dependend coupling constant) and 
arising because of the DR method [20], after the       
completion of the IRMR program, also become terms of order $\epsilon$ and      
therefore they vanish in the $\epsilon \rightarrow 0^+$ limit.                 
Obviously, the same is valid for all other finite terms in the expansion (3.7).
 
Absolutely in the same way should be done IRMR program in the ghost and        
quark-ghost sectors which through the corresponding ST identity provide        
information about the quark-gluon vertex function at zero momentum transfer    
relevant for the investigation of the DIR region in QCD. The corresponding convergent conditions of the ghost self-energy and ST identity could be 
also derived. Together with the quark SD convergence condition (4.16), they    
govern the concrete $\epsilon$-dependence of the IRMR constants which in      
general remain arbitrary. These three independent convergent conditions are    
the basis for removing from the theory all the NP IR divergences on a general  
ground and in self-consistent way. There exist two more independent 
convergence conditions including IRMR constants for three and four-gluon
proper vertices, of course.

\section{Quark confinement and DCSB}

Today there are no doubts left that the dynamical mechanisms
of the important NP quantum phenomena such as quark confinement   
and dynamical chiral symmetry breakdown (DCSB) are closely
related to the complicated topologically and dynamically nontrivial structure  
of the QCD NP vacuum [24-26]. On the other hand, it also becomes clear that    
the NP IR dynamical singularities, closely related to the nontrivial vacuum
structure, play an important role in the large distance behaviour
of QCD [27,28]. For this reason, any correct NP model of quark confinement and 
DCSB (closely related to each other) necessarily turns out to be a  
model of the true QCD vacuum and the other way around.

Let us now write down the final system of equations consisting of the quark SD 
equation and corresponding ST identity for the most interesting case of 4D INP 
QCD, namely
 
\begin{eqnarray}
S^{-1} (p) &=& S_0^{-1} (p)+ \bar {\mu}^2  \Gamma_\mu(p,0) S(p) \gamma_\mu, \nonumber \\
\Gamma_\mu(p,0) &=& id_\mu S^{-1}(p) - S(p) \Gamma_\mu(p,0) S^{-1}(p).
\end{eqnarray}
For simplicity here we removed "bars" from definitions of the IR finite Green's
functions and retaining it only for the NP mass scale parameter in order to    
distingush it from the initial mass scale parameter. The IR renormalized NP    
mass $\bar \mu^2$ includes all finite constants.                               
This system of equations corresponds to the simplest NP IR singularity possible
in 4D QCD, the so-called ``$q^{-4}$'' singularity which leads to linearly      
rising potential between heavy quarks within the Wilson loop approach [23].    
In order to get the first        
equation in this system, it is necessary to substitute the distribution (3.8)  
into the quark SD equation (4.6). The next step is to perform IRMR program how 
it was described in preceding section coming thus to the quark SD eqution      
(4.17)-(4.18) which in this case is nothing else but the first equation of the 
system (5.1). It is a general feature of all simplest NP IR singularities    
($k=0$) that corresponding arbitrary function $f(q^2)$ in Eq. (3.1) is to be   
replased by the first constant term of its Taylor expansion (3.2). So the      
vacuum in  
this case is labelled as $\lambda =-2, \ k=0$. Not loosing generality, we also 
put the quark wave function IRMR constant $Z_2(\epsilon)=1$, i.e., this      
system of equations corresponds to the IR finite from the very beginning quark 
propagator. For the IR vanishing (as $\epsilon \rightarrow 0^+$) quark         
propagator the inhomogeneous term (free quark propagator) in the quark SD      
equation (4.17) should be replaced by                                          
$\bar m_0 = Z_2(\epsilon) m_0 (\epsilon)$.   
    
In the same way should be evaluated ST identity in order to get the information
about quark-gluon vertex at zero momentum transfer [29]. It contains ghost contributions which should be taken into account with the help of the corresponding
SD equation for the ghost self-energy. In covariant gauge QCD, by contributing 
only into the closed loops, nothing should explicitly depend on ghost degrees  
of freedom. However, the ghost-quark sector contains very important piece of   
information about quark degrees of freedom themselves. Precisely this information has been taken into account in the second term of the ST identity which is  
nothing else but the second equation in the system (5.1). The first term presents standard contribution, where $d_{\mu} = d / dp_{\mu}$, by definition.   
In noncovariant gauge QCD ghosts do not exist but instead gauge dependent singularities appear there [1,30]. How to separate them from the dynamical NP IR singulatities is not a simple task. Anyway, though the corresponding ST identity 
will be standard, nevertheless, the corresponding quark SD equation will be    
much more complicated than in Eq. (5.1).

\subsection{Solution}

Thus our final system of equations to be solved is the system (5.1). Euclidean 
versions of the chosen parametrization for the free and full quark propagators 
are as follows:

\begin{equation}
S_0^{-1}(p)= i( \hat p + m_0)
\end{equation}
and

\begin{equation}
iS(p)= \hat p A(p^2) - B (p^2),
\end{equation}
respectively. Its inverse is

\begin{equation}
iS^{-1}(p)= \hat p \overline A(p^2) +  \overline B (p^2),
\end{equation}
where

\begin{eqnarray}
\overline A(p^2) &=& A(p^2) E^{-1}(p^2), \nonumber\\
\overline B(p^2) &=& B(p^2) E^{-1}(p^2), \nonumber\\
E(p^2) &=& p^2 A^2(p^2) + B^2(p^2).
\end{eqnarray}
In order to solve the ST identity (the second in the system (5.1)), the simplest way is to decompose the vertex at zero momentum transfer as a sum of its four
independent form factors. Introducing now the dimensionless variables and functions

\begin{equation}
A(p^2) = \bar \mu^{-2} A(x), \qquad B(p^2) = \bar \mu^{-1} B(x),  \qquad
 x = p^2/{\bar \mu^2},                                                         
\end{equation}
and after doing some tedious algebra of $\gamma$ matricies with a Euclidean    
signature in four dimentions, one finally obtains

\begin{eqnarray}
x A' &=& - (2 + x) A - 1 - m_0 B, \nonumber\\
2B B' &=& -3 A^2  + 2 ( m_0 A - B)B,
\end{eqnarray}
 where $A \equiv A(x)$, $B \equiv B(x)$, and
the prime denotes the derivative with
respect to the Euclidean dimensionless momentum variable $x$. Obviuosly we
retain the same notation for the dimensionless current quark    
mass, i. e. $m_0 / \bar \mu \rightarrow m_0$.
The system (5.1) and consequently (5.2) for the first time has been obtained   
in Ref. [31].

\subsection{Quark confinement}

 The exact solution of the system (5.7) for the dynamically generated quark    
mass function is

\begin{equation}
 B^2(c, x; m_0) =  \exp(- 2x)
\int^c_x \exp(2x') \tilde{\nu}(x')\, dx' ,
\end{equation}
where $c$ is the constant of integration and

\begin{equation}
\tilde{\nu} (x) = 3 A^2(x) +2 A(x) \nu(x)
\end{equation}
with

\begin{equation}
\nu (x) = - m_0 B(x) = xA'(x) + (2 + x)A(x) + 1.
\end{equation}
Then the equation for determining the $A(x)$ function becomes

\begin{equation}
{d\nu^2 (x) \over dx}+ 2\nu^2(x)= -3 A^2(x) m^2_0 - 2 A(x) \nu(x) m_0^2.
\end{equation}
Evidently, it is possible to develop the calculation schemes in different modifications which give the solution of this system step by step 
in powers of the light current quark masses as well as in the inverse powers of
the heavy quark masses.

The important observation, however, is that                               
the formal exact solution (5.8) exhibits the algebraic branch point
at $x=c$ which completely $excludes \ the \ pole-type \ singularity$           
at any finite point on the real axis in the $x$-complex plane     
whatever solution for the $A(x)$ function might be. 
Thus the solution cannot be presented as the expression      
having finally the pole-type singularity at any finite point $p^2 = - m^2$     
(Euclidean signature), i.e.,  

\begin{equation}
S(p) \neq { const \over \hat p + m},
\end{equation}
certainly satisfying thereby the first necessary condition of
quark confinement formulated at the fundamental quark level as the absense of  
the pole-type singularities in the quark propagator [32].                      
These unphysical singularities (branch points at $x=c$ and at infinity) are    
caused by the inevitable ghost contributions in the covariant gauges. 

In order to prove (5.12), let us assume the opposite, i.e., that the quark propagator within our approach, nevetheless, may have pole-type singularities. In  
terms of the quark dimensionless form factors, defined in Eq. (5.6), this means
that in the neighborhood of a possible pole at $x= -m^2$ (Euclidean signature),
they can be presented as follows: 

\begin{eqnarray}
A(x) &=& {1 \over (x + m^2)^{\alpha} } \tilde{A}(x), \nonumber\\
B(x) &=& {1 \over (x + m^2)^{\beta} } \tilde{B}(x),
\end{eqnarray}
where $\tilde{A}(x)$ and $\tilde{B}(x)$ with their first derivatives (at least)
are regular at pole. The exponents ${\alpha}$ and ${\beta}$ in general are arbitrary with $Re {\alpha}, {\beta} \geq 0$. However, substituting these          
expansions into the initial system of equations (5.7) and analysing it in the  
neigborhood of the pole, one can immediately conclude in that the self-
consistent system of equations for quantities with tilde corresponds only to  

\begin{equation}
\alpha = \beta =0, 
\end{equation}
i.e., our sytem of equations (5.7) does not admit the pole type singularities
in the quark propagator.

The second sufficient condition of confinement formulated at hadronic level as 
the existence of the $discrete$ spectrum only (no continuum spectrum) in the   
bound-state problems [21] is obviously beyond the scope of this paper.

\subsection{Chiral limit}

In the chiral limit ($m_0 = 0$) the system (5.7) can be solved exactly. The    
exact solution for the $A(x)$ function is

\begin{equation}
A(x) =  {1 \over x^2} \left\{ 1 - x - \exp(-x) \right\}.
\end{equation}
For the dynamically generated quark mass function $B(x)$ the exact solution is 

\begin{equation}
 B^2(x_0, x) = 3 \exp(- 2x)
\int^{x_0}_x \exp(2x') A^2(x')\, dx' ,
\end{equation}
where $x_0 = p_0^2 / \bar \mu^2$ is an arbitrary constant of integration. 
The solutions (5.15) and (5.16) for the $A(x)$ function and for the
dynamically generated quark mass function $B(x_0,x)$, respectively, 
are not entire functions. Functions $A(x)$
and $B(x_0,x)$ have removable singularities at zero, i.e., they are finite at  
zero points. In addition, dynamically generated quark mass function $B(x_0,x)$
also has algebraic branch points at $x=x_0$ and at infinity (at fixed $x_0$) due to ghosts as was mentioned above.                                            

The solution for the $A(x)$ is regular at zero and automatically has a correct 
behavior at infinity (it asymptotically approches the free propagator at infinity). In order to reproduce a correct behavior at infinity ($x \rightarrow      
\infty$) of the dynamically generated quark mass function $B^2(x)$, it is      
necessary to simultaneously pass to the limit ($x_0 \rightarrow \infty$ in     
Eq. (5.16), so it identically vanishes in this limit in accordance with the vanishing current light quark mass in the chiral limit. Obviously, we have to keep
the constant of integration $x_0$ arbitrary but finite in order to obtain a regular at zero point solution. The problem is that if $x_0 = \infty$ then the 
solution (5.16) does not exist at all at any finite $x$, in particular at      
$x=0$. Thus even at the fundamental quark level   
the finite constant of integration $x_0$ provides an exact criterion 
how to separate soft momenta region $x \leq x_0$, where the INP dynamics takes
place from hard momenta region $x > x_0$, where PT dynamics celebrates. 

Concluding, let us note that the singular at zero point exact solutions to the 
system (5.7) in the chiral limit also exist. It is easy to check that the exact
solution

\begin{equation}
A(x) =  {1 \over x^2} \left\{ 1 - x  \right\}
\end{equation}
automatically satisfies the system (5.7). Substituting it into the Eq. (5.16), 
one obtains the exact solution for the dynamically generated quark mass        
function which is singular at zero point.

\subsection{DCSB}

From the initial system of coupled differential equations (5.7) it is easy to  
see that the system                                                            
         
$allows \ a  \ chiral \ symmetry \ breaking \ solution \ only$
                                                                               
\begin{equation}  
m_0 = 0, \quad A(x) \ne 0, \ B(x) \ne 0                               
\end{equation}
and                                                                            
    
$forbids \ a \ chiral \ symmetry \ preserving \ solution$                 

\begin{equation}
m_0 = B(x) = 0, \quad A(x) \ne 0.
\end{equation}
Thus any nontrivial solutions automatically break the $\gamma_5$ invariance of
the quark propagator                                                          
 
\begin{equation}
\{ \gamma_5, S^{-1} (p) \} = -  i\gamma_5 2 \overline B(p^2) \neq 0, 
\end{equation}
and they therefore $certainly$ lead to the  spontaneous breakdown of chiral symmetry ($m_0 = 0, \ \overline{B}(x) \ne 0$,  dynamical quark mass generation). 
In all previous investigations a chiral symmetry preserving solution (5.18)   
always exists. We do not distinguish between $B(x)$ and $\overline{B}(x)$      
calling both the dynamically generated quark mass functions for simplicity.
Thus the measure of DCSB at the fundamental quark level is always twice of it.

\section{Discussion and Conclusions}

In summary, a few points are worth reemphasizing. The first is that the only   
place where the most important problem of theoretical physics --          
confinement (together with other NP effects) can be solved is the SD system of 
dynamical equations of motion since it contains the full dynamical information 
on QCD. To solve this system means to solve QCD itself and vice versa.         
The second is that AF clearly indicates the existence of the NP phase in QCD. 
 We propose how it   
can be exactly defined in the YM theory by explicitly introducing the NP mass  
scale parameter responsible for all NP effects in QCD within our approach,     
Eqs. (2.2) and (2.3).  In this way the NP phase becomes the TNP one,           
Eqs. (2.5)-(2.8). In the full QCD it is described by the quark
SD equation (4.2)-(4.4), where explicit gauge dependence is already shifted    
to the PT part. The standard UVRM program is needed 
to render it finite. The third point is that the only place where nontrivial   
dynamics can be introduced is the DIR region since the PT structure of QCD is  
controlled by AF. The DIR region is dominated by self-interaction of massless  
gluons in the QCD vacuum which becomes strongly singular in the DIR limit 
leading thus to the enhancement of zero momentum modes in the QCD NP vacuum.
We define severe NP IR singularities as those being more singular in the DIR   
limit than the exact singularity of the free gluon propagator which thus       
additionaly separates the PT phase from the TNP one. In this way the INP phase 
in the YM theory has emerged.                                                  
                      
Our quantum, dynamical ZMME model of the QCD true ground state is based on the 
existence and importance of such kind of the NP excitations and fluctuations   
of gluon field configurations there which are due to self-interaction of       
massless gluons only without explicit involving some extra degrees of freedom. 
They   
are to be summarized by the INP part of the full gluon propagator and are to be
effectively correctly described by its behavior in the DIR domain. As was   
explained above, these types of gluon field configurations necessary           
correspond to the strongly singular in the DIR limit gluon propagator.  
However, we argue that severe NP IR singularities can be put under control 
within DT, complemented by the DR method. We have explicitly shown that in     
terms of the IR regularization parameter $\epsilon$, any NP IR singularity is  
always a simple pole, $1/ \epsilon$ (see Eq. (3.7)). What will be drastically  
different for different NP IR
singularities is the residue at pole which means different solutions  
for the whole system of SD equations. Different solutions in turn    
means different vacua. They are labelled by two numbers: $k$ which is always 
positive integer and the exponent $\lambda$. The former determines the residue 
at pole. The latter determines the power of the NP IR singularity 
with respect to gluon momentum itself. Also the dimension $D$ of the Euclidean
space should be taken into account. At given number of $k$ one has the finite 
number of almost "degenerate" vacua depending on the chosen numerical value    
for $D$. At given number of $D$ one has completely different vacua which number
depends on the exponent $\lambda$ and $k$. Thus a highly nontrivial dynamical  
and topological structure of the QCD NP vacuum appears within our approach
(Table I).

Because of the severe NP IR divergences, all the Green's functions should be   
regularized with the help of the above-mentioned IR regularization parameter,  
introduced by the DR method. Since the singularity is always a simple pole, 
this makes it possible to perform IRMR program in order   
to remove all the NP IR divergences from the theory on a general ground and in 
self-consistent way. The most remarkable feature of this program is that   
the PT contributions to the corresponding SD equations vanish since they       
become of oder $\epsilon$ as $\epsilon \rightarrow 0^+$ because of the corresponding convergence conditions. They are key objects of our IRMR program. 
In this way one obtains the IR renormalized system of the              
corresponding SD equations which do not explicitly depend on a gauge fixing    
parameter (see for example Eqs. (4.17)-(4.18)). Here we would like to bring    
attention of the reader to the important observation. Because of the correct   
treatment of the NP IR singularities within DT, complemented by the DR method, 
all the system of SD equations in the INP phase is UV finite, i.e., it is free 
from UV divergences.      

For the most interesting case of 4D INP QCD the closed set of SD equations in
the quark sector is shown in Eq. (5.1) which is obviously manifestly gauge     
invariant. This system of equations corresponds to the simplest NP IR          
singularity possible in 4D QCD. The corresponding formal exact        
solution (5.8) for the dynamically generated quark mass function is regular    
at zero. It is essentially NP since it cannot be obtained by PT expansion.     
 Moreover, it is of confinement type, i.e., it   
cannot have pole-type singularities and it certainly requires DCSB since a     
chiral symmetry preserving solution is forbidden (see Eqs. (5.17)-(5.18)).     
They also have a correct PT limit at $\bar \mu^2 \rightarrow 0$ which is
equivalent to $x, x_0 \rightarrow \infty$. For example, the exact solution     
(5.14) for the dynamically generated qaurk mass function identically vanishes 
as it should be in the chiral limit while quark wave function form factor behaves like free quark propagator, i.e., $A(x) \sim - x^{-1}, \ x \rightarrow \infty$ (see Eq. (5.13)).       

Here it is worthwhile to make a few remarks.
The correct treatment of a such strong singularity (3.8)
within DT, complemented by the DR method, enables us to extract the required   
class of test functions in the IR renormalized quark SD equation               
(4.17)-(4.18). The test functions do consist of the quark propagator and       
the corresponding quark-gluon
vertex function. By the renormalization program we have
found the regular solutions for the quark propagator. 
For that very reason the relation (3.8) is justified, it is
multiplied by the appropriate smooth test functions [19]. Moreover, we have    
established the space in which our generilized functions are continuous 
linear functionals. It is a linear topological space, denoted as $K(c)$ (in    
the chiral limit $K(x_0)$), consisting  
of infinitely differentiable functions having compact supports, i.e., test functions vanish outside the interval $x \leq c$ (in the chiral limt $x \leq x_0$).
It is well known that this space can be identified with a complete countably   
normed space which is at the same time a linear metric space, and the topolgy 
defined by the metric is equivavalent to the original topology [19]. Thus the  
above and below-mentioned subtractions of all kind of the PT contributions
become not only physically well justified but mathematically (by DT) as well.  

In the chiral limit the system of equations (5.1) is exactly solvable (see     
Eqs. (5.15) and (5.16)). Chiral 4D INP QCD depends only on the constant of     
integration $x_0 = p_0^2 / \bar \mu^2$ of the quark SD equation and the NP     
scale $\bar \mu$ itself    
(in the general, nonchiral case the dependence on the current quark masses will
appear, of course). In order to determine them a physically well-motivated     
scale-setting scheme is needed since $\bar \mu$ determines the physical scale  
of INP dynamics within our approach, so it is always finite, i.e., it cannot
be arbitrary large. This has been already done in Refs. [14,15]. Our numerical 
results turned out to be in fair agreement with phenomenological values of the 
chiral condensate, dynamically generated quark masses and the pion decay       
constant in the chiral limit. Moreover, the topology of a chiral QCD NP vacuum 
is also in very good agreement with our values for INP vacuum energy density,
gluon condensate, topological susceptibility, etc. [33,34].   

At the very beginning of QCD it was expressed a general idea [27,28] that the 
quantum excitations in QCD vacuum of the IR degrees of freedom because of      
self-interaction of massless gluons only made it possible to understand        
confinement, DCSB and other NP effects. In other words, the importance of the
DIR structure of the true QCD vacuum has been emphasized as well as its        
relevance to quark confinement, DCSB, etc., and other way around. 
This development was stopped by the wide-spread wrong opinion that
severe NP IR singularities cannot be put under control. We have explicitly     
shown that correct mathematical theory of quantum YM physical theory is the    
theory of distributions (theory of generilized functions) [19], complemented by
the DR method [20]. They provide a correct treatment of these severe singularities without any problems. Thus we come back to the old idea but on a new basis 
that is why it becomes new ("new is well-forgotten old").

The correct treatment of the severe NP IR singularities allows one to          
discover a new phase in QCD, the INP one. In fact, QCD was consequently decomposed as follows:                                       

\begin{center}
 QCD $\longrightarrow$ (TNP \ and \ PT) $\longrightarrow$ (TNP + DIR \ and \ PT) $\longrightarrow$ (INP \ and \ PT) QCD.
\end{center}
Thus it consists of the two independent theories.
INP QCD is completely decoupled from PT QCD though behind both        
theories is the same main dynamical
effect in QCD, namely self-interaction of massless gluons in the QCD vacuum.
It leads to the explanation of all NP effects (confinement, DCSB,   
etc.) in the framework of INP QCD and to AF in the framework of PT QCD.    
In order to explain all the nontrivial dynamics of QCD we need no more extra   
degrees of freedom, though, of course, they can exist in the QCD vacuum.
INP and PT QCD are two different theories with different physical observables  
to calculate and with different characteristic fundamental scales. 
The SD system of equations in PT QCD explicitly depends on gauge fixing parameter. It is UV divergent theory and therefore needs nontrivial UVMR program to   
render it finite. After performing this program one may forget about it if somebody wants to concentrate his efforts in investigating low energy QCD.
        
Contrary to this, INP QCD is a manifestly gauge invariant theory, i.e., only   
transfer (physical) degrees of freedom of gauge bosons are relevant for INP    
QCD. It is free from UV divergences as well as it is regular at the    
origin at the fundamental (microscopic) quark-gluon level. However, at the hadronic (macroscopic) level this theory is also finite because of the above-mentioned structure of unphysical singularities of the solutions to the quark SD 
equation. Within INP QCD this implies that any physical observables (for       
example, the pion decay constant, hadron masses, etc.) should be     
determined by such $S$-matrix elements (correlation functions) from which all  
types of the PT contributions should be subtracted, by definition.  Let us     
remind  
the reader that many important quantities in QCD such as gluon and quark       
condensates, topological susceptibility, etc. are already defined in this    
way. This does not mean that Lorentz covariance is violated in this cases as   
well as in the case of INP QCD needed subtractions.                            
In INP QCD the subtraction point is precisely given by the corresponding  
constant of integration uniquely separating DIR region from the PT one, i.e.,  
it has a clear physical meaning and it cannot be arbitrary large, of course.   
On the other hand, this subtraction obviously 
makes it possible to cancel the dependence of physical observables on the      
above-mentioned inevitable unphysical singularities. Thus within INP QCD developed here in order to calculate any physical observable the only subtraction    
which is needed to be done is the subtraction of the integration over the PT   
region in the corresponding correlation function since all other ones have     
been already automatically done within our INP solutions.                      
                  
Let us also underline that such kind 
of subtractions are inevitable  for the sake of self-consistency as well. The  
problem is that in low energy QCD there exist
relations between different correlation functions. For example, the famous     
Witten-Veneziano (WV) formula [35] (see also Ref. [36]) relates the pion decay
constant and mass of the $\eta'$ meson to the topological susceptibility,   
and many similar others. Defining topological susceptibility by subtracting all
kind of the PT contributions it would be not self-consistent to retain them in 
the 
correlation functions defining the pion decay constant and mass of the $\eta'$ 
meson. Anyway, let us emphasize once more that INP QCD requires the above-discussed subtractions of all kind of the PT contributions, by definition. This     
guaruntees that none of the inevitable unphysical singularities mentioned above
will affect physical observables calculated within this theory.                
            
Thus in INP QCD everything is confined to finite volumes at the hadronic 
(macroscopic) level, though at the fundamental (microscopic) quark-gluon level 
our solutions formally are valid in the whole energy-momentum range. This      
theory allows one to calculate properties of         
observable hadrons starting from unobservable quark-gluon degrees of freedom,  
i.e., from underlying fundamental theory. This means that it makes it possible 
to calculate hadron properties from first principles. Unlike lattice QCD which
suffers from conceptual and many technical problems, in INP QCD all integrals  
are finite and they can be calculated to any requested accuracy.               
In INP QCD all numerical values of any physical  
observables will finally depend only on the characteristic mass scale parameter
(mass gap) and not on coupling constant like in phenomenological or PT QCD. 
This in agreement with the phenomenon of "dimensional transmutation" [1,12]   
which occurs whenever a massless theory acquires masses dynamically. It is a   
general feature of spontaneous symmetry breaking in field theories.

It is almost obvious that the correct theory of low energy QCD is precisely   
INP QCD and therefore it is a true theory of nuclear force as well.            
Having a finite mass gap in the sence of Jaffe and Witten (JW) [37],
it naturally explains why nuclear force is short-ranged despite gluons 
remain massless. In other words, INP QCD is essentially short-ranged           
fundamental theory. At the same time, the severe NP IR singularities,          
summarized by
the DIR structure of the full gluon propagator within INP QCD, lead to the     
so-called "IR slavery" which can be equivalently referred to as strong coupling
regime [1,27,28,38,39]. Thus nuclear force is strong due to precisely IR       
singular interaction of massless gluons. These properties of INP QCD do not    
depend on power of NP IR singularities, i.e., at any 
NP IR singularity these properties hold. It was not accindental that the     
explanation precisely of these properties of QCD was put on the first place by 
JW in Ref. [37]. In our opinion this is so indeed, since without explanation   
of these properties (though qualitatively) it is impossible to go further. 
In addition, exploring only simplest NP IR singularity possible in 4D QCD      
vacuum, we have explicitly shown that "IR slavery" implies quark confinement as
well as DCSB, indeed.                                                      

Concluding, it is worth reemphasizing once more that in INP QCD all
physical observables are to be expressed in terms of a mass gap $\Delta > 0$,  
i.e., in INP QCD there are no physical states in the interval $(0, \Delta)$   
in complete agreement with JW definition [37].
In addition, it has quark confinement as well as chiral symmetry breaking
in the sense explained by JW in the above-mentioned  Ref. [37], and which in   
turn are in complete agreement with our definitions. Thus by defining TNP QCD 
we introduce the mass gap in the JW sense since the subtraction (2.2) evidently
can be equivalently presented as follows:                                      

\begin{equation}
d^{TNP} (q^2, \Delta^2) = \Delta^2 f(q^2, \Delta^2),
\end{equation}
where the function of the correspondinf dimension  $f(q^2, \Delta^2)$ has a finite limit as $\Delta^2 \rightarrow 0$.
By going further in order to specify
the DIR structure of the theory we define INP QCD which becomes therefore      
manifestly gauge invariant.

In conclusion, let us note that there is nothing wrong with such strong        
singular behavior of the full gluon propagator in the DIR domain. The 
claster property of the Wightman functions fails in QCD [40] allowing thus for 
such behavior.  
%\acknowledgments

The author would like to thank H. Georgi, V.I. Zakharov, H. Fried,             
P. Forg\'acs, M. Faber, A. Ivanov and Gy. Kluge for useful correspondence,     
discussions and remarks.

\appendix
\section{The simplest NP IR singularities in QCD}

From the general discussion of the NP IR singularities possible in nD QCD      
(see section 3 above and Table I below), it
follows that the simplest ones always correspond to $k=0$, i.e.,               
$\lambda = - n/2$. As it follows from Eqs. (3.4) and (3.5), the residue at pole
in this case is  
always delta function of the corresponding dimension. It is rather easy to explicitly show that matrix form of the corresponding system of equations in nD momentum space with Euclidean signature for the IR regularized (finite) quark propagator is always given by the system (5.1), i.e., it does not explicitly depend
on $n$. However, the explicit dependence will appear after performing the algebra of the $\gamma$ matrices in nD Euclidean space. The corresponding system of 
equations for the quark propagator form factors becomes    

\begin{eqnarray}
x A' &=& - \Bigl( {n \over 2} + x \Bigr) A - 1 - m_0 B, \nonumber\\
2B B' &=& - (n-1) A^2  + 2 ( m_0 A - B)B,
\end{eqnarray}
 where, let us remind, $A \equiv A(x)$, $B \equiv B(x)$, and
the prime denotes the derivative with
respect to the Euclidean dimensionless momentum variable $x$. Obviuosly we
retain the same notation for the dimensionless current quark    
mass, i. e. $m_0 / \bar \mu \rightarrow m_0$. This system coincides with       
system (5.7) for $n=4$, as it should be. 

Similar to the solution (5.8), the exact solution for the dynamically generated
quark mass function in nD INP QCD is
 
\begin{equation}
 B^2(c_n, x; m_0) =  \exp(- 2x)
\int^{c_n}_x \exp(2x') \tilde{\nu}_n (x') dx' ,
\end{equation}
where $c_n$ are the corresponding constants of integration and

\begin{equation}
\tilde{\nu}_n (x) = (n-1) A^2(x) +2 A(x) \nu_n (x)
\end{equation}
with

\begin{equation}
\nu_n (x) = - m_0 B(x) = xA'(x) + \Bigl( {n \over 2} + x \Bigr) A(x) + 1.
\end{equation}
Then the equation for determining the $A(x)$ function becomes

\begin{equation}
{d\nu^2_n (x) \over dx}+ 2\nu^2_n(x)= -(n-1) A^2(x) m^2_0 - 2 A(x) \nu_n(x) m_0^2.
\end{equation}
Evidently, in this general case it is also possible to develop the calculation 
schemes in different modifications which give the solution of this system step 
by step in powers of the light current quark masses as well as in the inverse  
powers of the heavy quark masses.

Again, the important observation, however, is that                             
the formal exact solution (A2) exhibits the algebraic branch point
at $x=c_n$ which completely $excludes \ the \ pole-type \ singularity$         
at any finite point on the real axis in the $x$-complex plane     
whatever solution for the $A(x)$ function might be. 
Thus the solution cannot be presented as the expression      
having finally the pole-type singularity at any finite point $p^2 = - m^2$     
(Euclidean signature), i.e.,

\begin{equation}
S(p) \neq { const \over \hat p + m},
\end{equation}
certainly satisfying thereby the first necessary condition of
quark confinement formulated at the fundamental quark level as the absense of  
the pole-type singularities in the quark propagator [32].                      
These unphysical singularities (branch points at $x=c_n$ and at infinity) are  
caused by the inevitable ghost contributions in the covariant gauges. 

In the chiral limit ($m_0 = 0$) the system (A1) can be solved exactly. The    
exact solution for the $A(x)$ function is

\begin{equation}
A(x) = - (-x)^{-n/2} e^{-x} \gamma \Bigl( {n \over 2}, - x \Bigr).
\end{equation}
where $\gamma$ function is incomplete gamma function. For even number of dimension $n /2 = m'$, where $m'$ is integer, it is

\begin{equation}
\gamma (m', x) =  \gamma (1+ (m'-1), x) = (m'-1)! \Bigl( 1 - e^{-x}            
\sum_{m=0}^{m'-1} { x^m \over m!} \Bigr),
\end{equation}
otherwise it is determined by the infinite series.\footnote{Apparently, even   
for the odd number of dimensions the system (A1) remains valid though, as was  
mentioned above, the poles of the second order appear in this case.}   

For the dynamically generated quark mass function $B(x)$ the exact solution is 

\begin{equation}
 B^2(x_0, x) = (n-1) \exp(- 2x) \int^{x_0}_x \exp(2x') A^2(x') dx',
\end{equation}
where $x_0 = p_0^2 / \bar \mu^2$ is an arbitrary constant of integration, different for different $n$, of course. 

It is instructive to write down solutions for particular cases. The most interesting ones are 4D and 2D INP QCD. For 4D INP QCD the exact solutions
are explicitly shown in Eqs. (5.13) and (5.14). For 2D INP QCD the exact solutions are

\begin{equation}
A(x) = - {1 \over x} \left\{ 1 - \exp(-x) \right\}
\end{equation}
and 
\begin{equation}
 B^2(x_0, x) = \exp(- 2x) \int^{x_0}_x \exp(2x') A^2(x') dx',
\end{equation}
where $x_0 = p_0^2 / \bar \mu^2$ is again an arbitrary constant of integration.

Thus nD INP QCD confines quarks in the sense that the quark propagator has no  
poles and this does not depend on $n$ at all. Obviously, what is needed is only
self-interaction of massless gluons which is the main chracteristic of  
QCD in any dimensions. That is the spontaneous breakdown of chiral symmetry    
also does not depend on $n$ (from the general sysytem (A1) it follows that again a chiral symmetry preserving solution (5.16) is forbidden) is a result of the
above-mentioned main interection of QCD as well.     

Concluding, a few technical remarks are in order.
As was emphasized, the relevant system of equations in the matrix form (5.1)   
does not explicitly depend on number of dimensions $n$. This is true for the mass scale parameter which appears in this system. Indeed, in nD QCD the coupling
constant has dimension as follows: $g^2 = [\mu]^{4-n}$. On the other hand, from
expression (3.1) we also have a mass scale parameter $(\mu^2)^{- \lambda -1}$.
Thus for $\lambda = - (n/2)$ at $k=0$ (simplest singularities possible in nD   
QCD) the final combination of the characteristic  mass scale parameters is   
 
\begin{equation}
(\mu^2)^{- \lambda -1} g^2 = [\mu]^{n-2} [\mu]^{4-n}= [\mu]^2
\end{equation}
This unique mass scale parameter should be multiplied by the dimensionless     
constant, which depends, of course, on $n$, i.e.,

\begin{equation}
C_n \times \mu^2 \rightarrow \bar \mu_n^2 \rightarrow \bar \mu^2,   
\end{equation}
where we have finally omitted the explicit dependence on $n$ for simplicity but
it is always assumed, of course. For example, in 2D QCD the coupling constant  
itself has dimension of mass (which underlines its special status), so         
$\bar \mu$ should be understood as a coupling constant in this case with all 
finite numbers included [41].

\vfill

\eject

\vfill

\eject

\Large
\begin{table}
% Table 1
\caption{Classification of vacua within ZMME quantum model, $(q^2)^{\lambda}$}
\begin{center}
\renewcommand{\arraystretch}{1.5}
\begin{tabular}{lrrrr}
   & 1 & 2 &  3 & 4  \\
\hline
0 &$(q^2)^{-1/2}$  &  $(q^2)^{-1}$ & $(q^2)^{-3/2}$ & $(q^2)^{-2}$ \\
1 &$(q^2)^{-3/2}$  &  $(q^2)^{-2}$ & $(q^2)^{-5/2}$ & $(q^2)^{-3}$ \\    
2 & $(q^2)^{-5/2}$  &  $(q^2)^{-3}$ & $(q^2)^{-7/2}$ & $(q^2)^{-4}$ \\  
3 & $(q^2)^{-7/2}$  &  $(q^2)^{-4}$ & $(q^2)^{-9/2}$ & $(q^2)^{-5}$ \\ 
\end{tabular}
\renewcommand{\arraystretch}{1.0}
\end{center}
\end{table}

\vspace{5mm}

In this Table the main horisontal line is for dimensions of Euclidean space    
$D=n = 1, 2, 3, 4, ...$, while the vertical line is for $k=0, 1, 2, 3, ...$,
which determines the residue at pole.

The exponent is given as follows:

\begin{center}
$\lambda = - {D \over 2} - k$ 
\end{center}

\end{document}